\documentclass[%
 aip,
 twocolumn,
 amsmath,amssymb,
 reprint,%
]{revtex4-2}

\usepackage{graphicx}
\usepackage{dcolumn}
\usepackage{bm}
\usepackage{color}				
\usepackage{amsmath}

\newcommand{\code}[1]{\texttt{#1}}

\begin{document}

\preprint{AIP/123-QED}

\title{Recovering non-Maxwellian particle velocity distribution functions from collective Thomson-scattered spectra}

\author{B. C. Foo}
\affiliation{Princeton University, Princeton, NJ, USA}

\author{D. B. Schaeffer}%
\email{derek.schaeffer@ucla.edu}
\affiliation{University of California Los Angeles, Los Angeles, CA, USA}%

\author{P. V. Heuer}
\affiliation{University of Rochester Laboratory for Laser Energetics, Rochester, NY, USA}%

\date{\today}

\begin{abstract}
Collective optical Thomson scattering (TS) is a diagnostic commonly used to characterize plasma parameters. These parameters are typically extracted by a fitting algorithm that minimizes the difference between a measured scattered spectrum and an analytic spectrum calculated from the velocity distribution function (VDF) of the plasma. However, most existing TS analysis algorithms assume the VDFs are Maxwellian, and applying an algorithm which makes this assumption does not accurately extract the plasma parameters of a non-Maxwellian plasma due to the effect of non-Maxwellian deviations on the TS spectra. We present new open-source numerical tools for forward modeling analytic spectra from arbitrary VDFs, and show that these tools are able to more accurately extract plasma parameters from synthetic TS spectra generated by non-Maxwellian VDFs compared to standard TS algorithms. Estimated posterior probability distributions of fits to synthetic spectra for a variety of example non-Maxwellian VDFs are used to determine uncertainties in the extracted plasma parameters, and show that correlations between parameters can significantly affect the accuracy of fits in plasmas with non-Maxwellian VDFs.

\end{abstract}


\maketitle

\section{\label{sec:intro}Introduction}

Thomson scattering (TS) refers to the scattering of electromagnetic radiation by a collection of many charged particles, such as a plasma\cite{sheffield_2011, Froula2007quenching}. Optical TS, in which the probing radiation consists of optical wavelengths, is a widely-used \textit{in situ}, non-perturbative diagnostic tool for characterizing plasmas, with applications ranging from laboratory astrophysics~\cite{Schaeffer_2017,schaeffer_direct_2019,Bruulsema2020,Morita2020} to fusion plasmas~\cite{glenzer1997thomson,Nielsen_2016,Turnbull2020,Bruulsema2022}. 

In collective TS, the parameters associated with the plasmas of interest cause incident optical radiation to be scattered by electron plasma waves (EPW) and ion acoustic waves (IAW). The fluctuations in the charge density produce TS spectral features\cite{sheffield_2011} which can be measured and analyzed in order to extract information about the particle velocity distribution functions (VDFs). Most TS analysis tools assume that the plasma is thermalized so that the electron and ion VDFs are both Maxwellian with corresponding parameters like the temperatures $T_e$, $T_j$ and the densities $n_e$, $n_j,$ of the electron and ion populations. These Maxwellian parameters can then be inferred from the spectrum by an algorithm that minimizes the difference between the measured scattered spectrum and an analytic spectrum generated from these Maxwellian VDFs. 

Non-Maxwellian VDFs have been directly observed with TS diagnostics in recent experiments on high-energy-density (HED) plasmas \cite{Henchen2019,Milder2020evolution, milder_measurements} and laser-driven collisionless shocks\cite{schaeffer_direct_2019,Yamazaki2022}. Indeed, deviations of the VDF from a Maxwellian can be crucial to understanding the plasma dynamics. Previous theoretical studies have discussed the form of the TS spectra generated from non-Maxwellian distributions such as a two-stream distribution\cite{Sakai2020,Sakai_2023} and a super-Gaussian distribution\cite{Zheng1997}, as well as how the resulting spectra deviate from their Maxwellian counterparts. Milder \textit{et al}. \cite{Milder2019} have also shown that fitting the TS spectrum produced by a super-Gaussian electron VDF with a Maxwellian model can give incorrect plasma parameters. The inaccurate fitting is due to how the non-Maxwellian deviations affect the strength of Landau damping at the location of the Thomson spectral peaks. Consequently, existing tools designed for Maxwellian VDFs are insufficient to analyze TS spectra from plasmas with non-Maxwellian VDFs. 

In this paper we investigate how non-Maxwellian VDFs impact TS spectra with the aid of two new numerical tools that we have developed\cite{foo_2023_code}. The first tool is a ``forward model" which computes the analytic TS spectrum from a set of arbitrary discretized particle VDFs, and the second is a ``fitting algorithm" which extracts non-Maxwellian plasma parameters from a TS spectrum by iteratively applying the non-Maxwellian forward model at different points in parameter space. These open-source tools expand on the work of Milder \textit{et al}. by enabling the study of arbitrary non-Maxwellian VDFs. 

For the fitting algorithm, we examine the use of two schemes to optimize the parameter space exploration. One scheme is differential evolution (DE), which attempts to continuously optimize a candidate solution by combining previous solutions. This scheme can search a large parameter space and is often computationally more efficient that ``brute force'' methods, but it is susceptible to uncertainties caused by correlations between parameters in the solution. The second scheme is a Monte Carlo Markov Chain (MCMC).  While generally computationally more expensive than DE, MCMC is well-suited for exploring very large parameter spaces that can be used to estimate the uncertainties in the best-fit parameters from the DE scheme.

In Sec.~\ref{sec:ts} we review the TS theory that forms the basis for our method. The numerical tools we developed to implement this routine are discussed in Sec.~\ref{sec:forward_models}-\ref{sec:fitting_algorithms}. We also describe a process for testing our method and comparing it with an open-source method that assumes Maxwellian VDFs, the results of which are presented in Sec.~\ref{sec:results}. Finally, in Sec.~\ref{sec:error_bars} we analyze the uncertainty and confidence associated with our fitting algorithm and discuss possible areas of improvement. Our conclusions are summarized in Sec.~\ref{sec:conclusions}.

\section{\label{sec:ts}Thomson Scattering Theory}

In this section we briefly review the theory behind Thomson scattering, following the approach of Froula \textit{et al}~\cite{sheffield_2011}. TS occurs when the absorption of an incident photon causes a charged particle to undergo acceleration, which then induces Larmor radiation as the charge emits a photon. In the charge's rest frame (the primed frame), the frequency $\omega'_s$ of the scattered photon is equal to the frequency $\omega'_i$ of the incident photon. In the (unprimed) lab frame, $\omega_s$ can be solved for by computing the primed frame solution and applying the appropriate Doppler shifts, which are a function of the particle velocity $\mathbf{v}$ as seen in the lab frame, as well as the incident and scattered wavevectors $\textbf{k}_i$ and $\textbf{k}_s$, respectively. If we define $\omega \equiv \omega_s - \omega_i$ and $\textbf{k} \equiv \textbf{k}_s - \textbf{k}_i$, the final result is shown to be
\begin{equation}
    \omega = \mathbf{k}\cdot\mathbf{v}.
    \label{eq:1particle_TS}
\end{equation}
In the case of many charges, monochromatic incident light is scattered into a spectrum of frequencies, which is determined by the velocities of all charged particles in the plasma. In that case, the scattered power density $P$ has the following proportionality:
\begin{equation}
    P(\textbf{k}_s, \omega_s)\propto \left(1 + \frac{2\omega}{\omega_i}\right)S(\textbf{k}, \omega).
    \label{eq:scattered_power}
\end{equation}
The factor $S(\textbf{k}, \omega)$ is the spectral density function, which contains the dependence on the velocities of the charged particles in the plasma. Note that when discussing TS forward models and fitting algorithms, the ``TS spectrum" being computed and fitted is either the spectral density function itself, or the scattered power as in Eq.~\ref{eq:scattered_power}, depending on the data being fit. For the purposes of this paper, the TS spectrum is the normalized scattered power unless otherwise specified. In general, the spectral density function can be written in terms of the normalized VDFs of the electrons and ions in the plasma:

\begin{multline}
    S(\textbf{k}, \omega) = \underbrace{\frac{2\pi}{k}\bigg |1 - \frac{\chi_e}{\epsilon}\bigg |^2f_{eo}(\omega / k)}_{\text{Electron component}}\\
    + \underbrace{\sum_j \frac{2\pi}{k}\frac{Z_j^2n_{j0}}{N}\bigg |\frac{\chi_j}{\epsilon}\bigg |^2f_{jo}(\omega / k)}_{\text{Ion component}}.
    \label{eq:spectral_density}
\end{multline}

Here $f_{eo}$ is the one-dimensional electron VDF in the direction of measurement and $f_{jo}$ are the ion VDFs, with $j$ indexing the ion species. $Z_j$ and $n_{j0}$ are the charge and density of ion species respectively, and $N$ is the combined density of all ions. The electron susceptibility $\chi_e$ and the ion susceptibilities $\chi_j$ are functions of $\mathbf{k}$ and $\omega$ given by
\begin{align}
    \chi_e(\mathbf{k}, \omega) &= \frac{4\pi e^2 n_{e0}}{m_ek^2}\int_{-\infty}^{\infty} d\textbf{v}\frac{\textbf{k}\cdot\partial f_{e0}/\partial \textbf{v}}{\omega - \textbf{k}\cdot\textbf{v}} 
    \label{eq:echi}
    \\
    \chi_{j}(\mathbf{k}, \omega) &= \frac{4\pi Z^2e^2 n_{j0}}{m_j k^2}\int_{-\infty}^{\infty} d\textbf{v}\frac{\textbf{k}\cdot\partial f_{j0}/\partial \textbf{v}}{\omega - \textbf{k}\cdot\textbf{v}},
    \label{eq:ichi}
\end{align}
where $n_{e0}$ is the electron density, $e$ is the electric charge, $m_e$ is the electron mass, and the integrals can be performed along a Landau contour which deviates from the real axis just enough to avoid the pole at $v = \omega/k.$ The longitudinal dielectric function $\epsilon$ is given by
\begin{equation}
    \epsilon = 1+\chi_e + \sum_j \chi_j.
    \label{eq:epsilon}
\end{equation}
In practice, the TS will not greatly perturb the photon frequency, so the TS spectrum will only be non-negligible in a small range of scattered frequencies around the incident frequency, $\omega_s\approx \omega_i.$ Therefore, $\mathbf{k}_s$ and also $\mathbf{k}$ will also not vary significantly. If we make the approximation that the direction of $\mathbf{k}$ is effectively fixed, then $\chi_e$ and $\chi_j$ will only be sensitive to the 1D projected VDF in the direction of $\mathbf{k}$ and Eqs.~\ref{eq:echi}-\ref{eq:ichi} can be rewritten as integrals over scalars.

TS can either be non-collective, meaning dominated by scattering off of individual, non-correlated electrons in the plasma, or collective, meaning, in unmagnetized plasmas, dominated by electron plasma waves (EPWs) and ion acoustic waves (IAWs) which propagate through the plasma. These regimes can be distinguished by the scattering parameter $\alpha$, defined in terms of the electron Debye length $\lambda_{De}$ as
\begin{equation}
    \alpha = \frac{1}{k\lambda_{De}}.
\end{equation}
If $\alpha\ll 1,$ then $\lambda\ll \lambda_{De}$ and the incident radiation effectively sees the electrons as free (unbound), leading to the non-collective regime. If $\alpha\gtrsim 1$, then the radiation sees the Debye-shielded charges and therefore the correlations between the motions of the electrons, leading to collective scattering.
The collective TS spectrum can be broken into two regimes. At high $|\omega|$, the heavier ions are unable to respond, so the electron component as labeled in Eq.~\ref{eq:spectral_density} dominates. At low $|\omega|$, both the electrons and ions are able to respond, but the ion component tends to dominate. The electron component is generally still non-trivial, which becomes important when applying a fitting algorithm. We will refer to the high $|\omega|$ regime as the EPW spectrum, as it is dominated by scattering from EPWs, and the low $|\omega|$ regime as the IAW spectrum, as it is dominated by scattering from IAWs. Due to the difference in frequency scales between the electron- and ion-dominated components, the EPW and IAW spectra are usually measured with separate spectrometers in experiments~\cite{Katz2012, schaeffer_direct_2019}, so treating them separately is justified. Additionally, the measurement of the EPW spectrum usually involves the placement of a notch filter to block out the low-$\omega$ portion of the spectrum as it is much brighter than the high-$\omega$ electron contributions. Our code takes this notch filter into account.

Because Eq.~\ref{eq:spectral_density} can be used to compute the TS spectrum from an arbitrary set of VDFs, in principle it may be possible to invert this process and infer the VDFs from a TS spectrum, assuming that the TS spectrum is a non-degenerate function of the VDFs. In practice this inversion can be unreliable even if the TS spectrum is nearly degenerate or if the VDFs are described by too many parameters.

This inversion process is known to work if the VDFs in question are Maxwellian\cite{sheffield_2011}, and most tools developed for TS analysis use this Maxwellian assumption. However, these tools cannot accurately reconstruct the VDFs when this assumption is violated, as described in following sections.

\subsection{\label{sec:vdfs}Example VDFs}
Although in theory we can compute a TS spectrum from arbitrary VDFs, the examples presented in the figures in this paper focus on a representative selection of non-Maxwellian VDF models relevant to plasma physics. These VDF models as well as their plasma parameters are discussed below.

Maxwellians are the most common VDFs in plasmas, as they describe plasmas which are in thermal equilibrium and can be derived from statistical mechanics and the Boltzmann distribution. A Maxwellian distribution takes the form
\begin{equation}
    f(v) = N\exp{\bigg(-\frac{m(v - v_D)^2}{2k_BT}\bigg)},
    \label{eq:Maxwellian_def}
\end{equation}
where $k_B$ is the Boltzmann constant and $N$ is some normalization factor. A unit-normalized Maxwellian has two parameters: the drift velocity $v_D$ and the temperature $T$. A VDF could also be composed of a linear combination of two or more Maxwellians which have different temperatures and/or drift velocities.\\

The kappa or generalized Lorentzian distribution is a non-Maxwellian distribution which is expected in plasmas where collisions are insufficient to thermalize the plasma, leaving more particles at high energy and forming suprathermal tails which deviate from Maxwellians. The integral-normalized kappa distribution takes the form
\begin{equation}
    f(v) = \frac{\kappa^{-\frac{3}{2}}}{2\pi w_{\kappa}^3}\frac{\Gamma(\kappa + 1)}{\Gamma(\kappa - \frac{1}{2})\Gamma(\frac{3}{2})}\bigg(1 + \frac{(v - v_D)^2}{\kappa w_\kappa^2}\bigg)^{-(\kappa + 1)},
    \label{eq:kappa_def}
\end{equation}
with $w_\kappa = \sqrt{(2\kappa - 3)k_B T / \kappa m}$ for particles of mass $m$. Here $v_D$ is the drift velocity, $\Gamma()$ is the Gamma function, and the spectral index $\kappa$ is a measure of the non-Maxwellian deviation\cite{Pierrard_2010}. Note that the usual notion of temperature does not apply to non-Maxwellian distributions as they are not thermalized, but we can still define an equivalent temperature $T$ in terms of the variance of the VDF:
\begin{equation}
    T\equiv C\int dv f(v) (v-v_D)^2,
\end{equation}
where $C$ is an appropriate normalization such that $T$ matches the usual temperature for Maxwellians, and $v_D$ is the mean drift of the VDF, defined as
\begin{equation}
    v_D \equiv \int dv f(v)v.
\end{equation}

In the limit $\kappa\to\infty$, the kappa distribution approaches a Maxwellian with the same drift velocity and equivalent temperature, while at $\kappa=3/2$, the distribution collapses and becomes undefined. Taking a linear combination of a hot kappa distribution and a cooler Maxwellian results in a core-halo distribution, which is commonly observed in the solar wind\cite{Pierrard2001}.

The super-Gaussian distribution is another common non-Maxwellian VDF model in HED plasmas and can be created by inverse bremsstrahlung heating\cite{Milder2019}. The general form looks similar to a Maxwellian or Gaussian distribution but with the power in the exponent as an additional free parameter. The 3D isotropic super-Gaussian distribution takes the form\cite{Milder2019}
\begin{equation}
    F(\mathbf{v}) = \frac{p}{4\pi v_p^3\Gamma(3/p)} \exp\left(-\left|\frac{\mathbf{v} - \mathbf{v}_B}{v_p}\right|^p\right),
    \label{eq:isotropic_supergaussian}
\end{equation}
with drift velocity $\mathbf{v}_D$ and additional parameters $x$ and $p$. When $p = 2$ this reduces to a Maxwellian (Eq.~\ref{eq:Maxwellian_def}) and $x$ becomes the temperature (up to a constant). For $p\neq 2$, the parameter $x$ remains linearly related to the temperature via a constant factor which depends on $p$. The corresponding 1D projection is given by
\begin{equation}
    f(v_x) = \int F(\mathbf{v})\;dv_ydv_z,
\end{equation}
which can be computed in terms of gamma functions. A related VDF which is simpler to express is the 1D super-Gaussian:
\begin{equation}
    f(v) = \frac{p}{2v_p\Gamma(1/p)}\exp\left(-\left|\frac{v - v_B}{v_p}\right|^p\right).
    \label{eq:1d_supergaussian}
\end{equation}

\begin{figure}
    \centering
    \includegraphics[width = \linewidth]{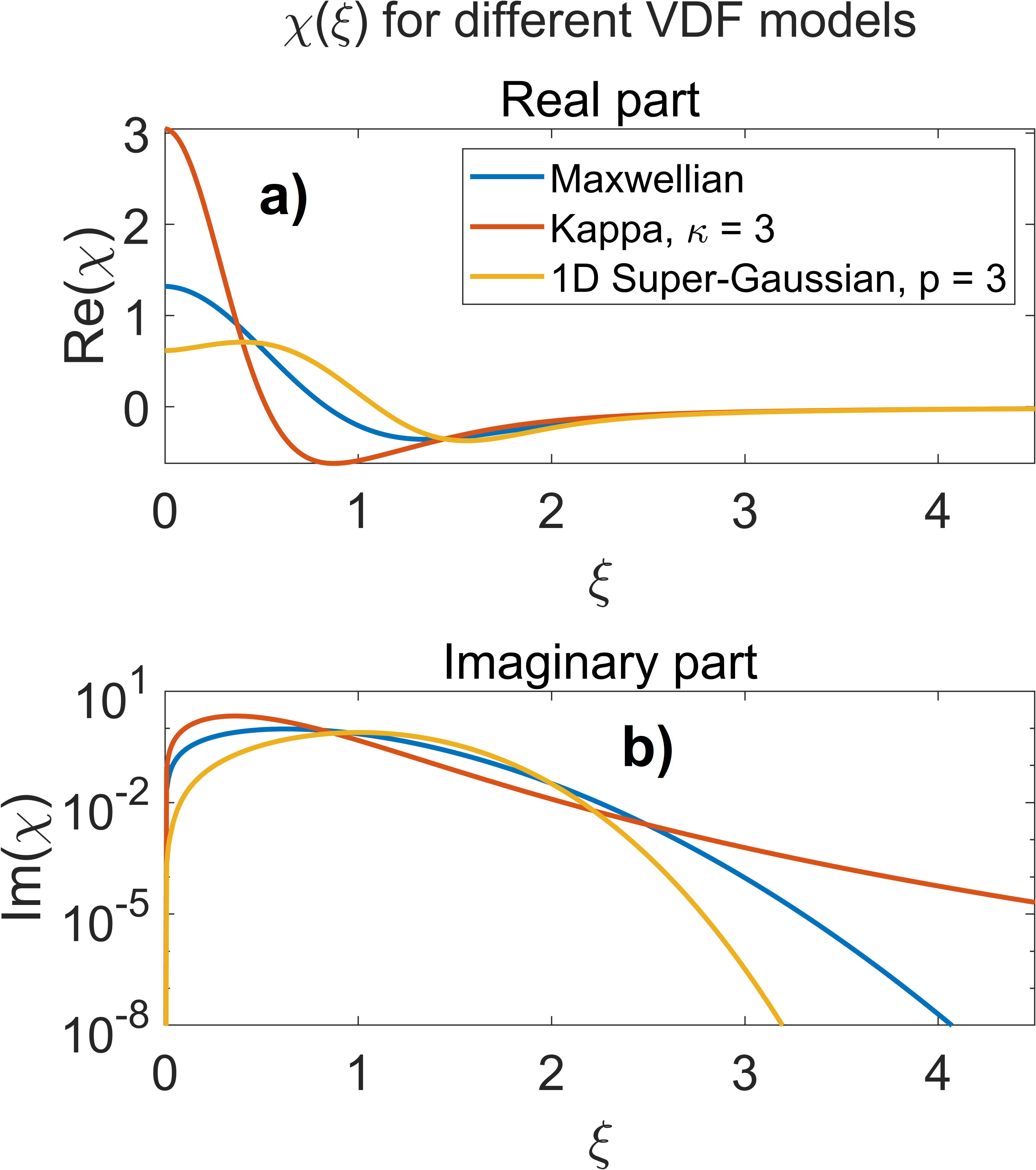}
    \caption{Comparison of $\chi(\xi)$ for three different VDF models: Maxwellian (blue), kappa ($\kappa = 3$, red), and 1D super-Gaussian ($p = 3$, yellow). The real part of $\chi$ is shown in a) and the imaginary part is shown in b). The imaginary part is shown on a semi-log plot to emphasize the distinction between $\chi$ for the different VDF models at high $\xi$.}
    \label{fig:chi_plots}
\end{figure}

\subsection{$\chi$ for different VDF models}
Recalling the definitions of the $\chi$ functions defined in Eqs.~\ref{eq:echi}-\ref{eq:ichi}, note that if we perform a change of variables $\mathbf{u} = \mathbf{v} / \left( v_{th}\sqrt{2}\right)$, $\xi = \omega / \left(k v_{th}\sqrt{2}\right)$ and operate in the regime where $k$ is nearly fixed, then we can write $\chi$ as a function of $\xi$, up to some constant factors determined by the plasma parameters:
\begin{align}
    \chi_e(\xi) &\propto \int_{-\infty}^{\infty} d\textbf{u}\frac{\partial f_{e0}/\partial \textbf{u}}{\xi - \textbf{u}},
    \\
    \chi_{j}(\xi) &\propto \int_{-\infty}^{\infty} d\textbf{u}\frac{\partial f_{j0}/\partial \textbf{u}}{\xi - \textbf{u}}.
\end{align}

These dimensionless integrals have the temperature dependence factored out, so they only depend on the overall form of the VDF and will differ for different VDF models. Fig.~\ref{fig:chi_plots} shows the dimensionless $\chi(\xi)$ integrals for three different VDF models for reference. It is helpful to note that the real part of $\chi$ represents the dispersion of an electrostatic wave at given $\omega$ through the plasma, while the the imaginary part represents the Landau damping on electrostatic waves.

\begin{figure*}
    \centering
    \includegraphics[width = 1\textwidth]{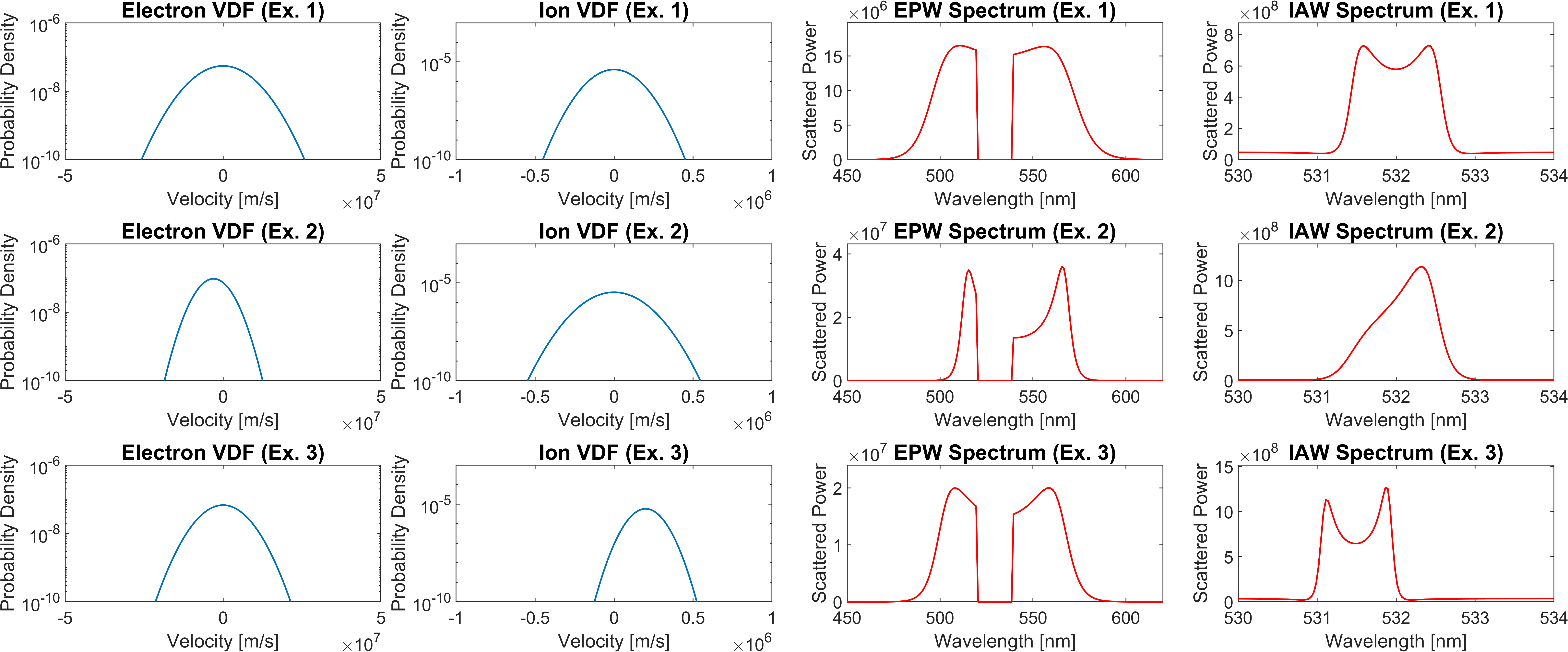}
    \caption{Normalized Maxwellian electron and ion (proton) VDFs, and the resulting normalized EPW and IAW spectra for three different sets of plasma parameters. The different examples are arranged by row. All examples assume an electron-proton plasma with a number density of $n = 4\times10^{18}\;\text{cm}^{-3}$ for each species. The incident laser wavelength is $532\;\text{nm}.$ The EPW spectra each have the wavelength range $[520, 540]\;\text{nm}$ excluded to represent a notch filter. The Maxwellian plasma parameters are indicated in Table~\ref{tab:maxwellian_parameters}, with the example numbers in the subtitles of each plot indicating which set of parameters it is associated with.
    All TS spectra were computed using the Maxwellian forward model from PlasmaPy\cite{PlasmaPyv1.0}. Each spectrum is normalized to have unit integral when computed in SI units, after accounting for the notch filter.}
    \label{fig:maxwellian_spectra}
\end{figure*}

The $\chi$ functions significantly impact the features of the TS spectrum. For instance, a larger imaginary part of $\chi$, corresponding to stronger Landau damping, is associated with broader spectral peaks \cite{Milder2019}. This can be seen in Fig.~\ref{fig:maxwellian_spectra}, which shows several analytic TS spectra from Maxwellian and drifting Maxwellian VDFs. The location of the EPW spectral peaks depends on the dispersion relation of the EPWs\cite{sheffield_2011}, which only weakly depends on the temperature at optical wavelengths (assuming the charge density is fixed). Therefore, increasing the temperature leaves the location of the spectral peaks essentially unchanged in velocity space, but increases the thermal velocity and therefore moves the spectral peaks to lower $\xi$, which corresponds to higher imaginary $\chi$ (for $\xi\gtrsim1$). Thus, we expect broader spectral peaks at higher temperatures, which is seen in the EPW TS spectra in Fig.~\ref{fig:maxwellian_spectra}.

\begin{table*}
    \centering
    \begin{tabular}{c|ccccccc}
    Ex. \# & $T_e$ [eV] & $v_e$ [m/s] & $T_i$ [eV] & $v_i$ [m/s] & $n_e$ [cm$^{-3}$]& $n_i$ [cm$^{-3}$] & Resolution ($\delta v_e$, $\delta v_i$) [m/s]\\
    \hline
    1 & 300 & 0 & 100 & 0 & $4\times 10^{18}$ & $4\times 10^{18}$ & $(2\times 10^5, 8\times 10^3)$\\

    2 & 100 & $-3\times 10^6$ & 150 & 0 & $4\times 10^{18}$ & $4\times 10^{18}$ & ---\\

    3 & 200 & 0 & 50 & $2\times 10^5$ & $4\times 10^{18}$ & $4\times 10^{18}$ & ---\\

    4 & 300 & 0 & (100, 200) & (0, 3)$\times 10^5$ & $4\times 10^{18}$ & $(1.6, 2.4)\times 10^{18}$ & $(2\times 10^5, 8\times 10^3)$\\

    5 & 300 & 0 & 100 & 0 & $4\times 10^{18}$ & $4\times 10^{18}$ & $(8\times 10^6, 4\times 10^5)$\\

    6 & 300 & 0 & 100 & 0 & $4\times 10^{18}$ & $4\times 10^{18}$ & $(8\times 10^4, 1.2\times 10^3)$\\
    
\end{tabular}
    \caption{Example Maxwellian plasma parameters used for calculating VDFs and associated TS spectra. In all cases except Example 4, both the electron and ion VDFs are defined by a single Maxwellian. In Example 4, the ion VDF is composed of two Maxwellians, the parameters of which are indicated by ordered pairs. For Examples 1, 4, 5, and 6, the discretized velocity spacing $\delta v_e$ and $\delta v_i$ for the electron and ion VDFs, respectively, are also shown for reference. These parameters affect the input of the arbitrary forward model. For the purposes of these example spectra, 500 velocity points are used for each VDF. The Resolution column is empty for Examples 2 and 3 as only the Maxwellian forward model, which does not take discretized VDFs as input, is applied to those examples. Note that Examples 5 and 6 have the same plasma parameters as Example 1, but different resolutions in the numerical computation of the TS spectra, as shown in Fig.~\ref{fig:cold_hot_VDFs}.}
    \label{tab:maxwellian_parameters}
\end{table*}

Non-Maxwellian deviations in the VDFs impact $\chi$ even if the temperature and density are held constant, which then affects the TS spectrum. A purely Maxwellian fitting algorithm can only vary the Maxwellian temperature and density, so when fitting a TS spectrum with non-Maxwellian deviations in $\chi$, the algorithm will attempt to compensate for those non-Maxwellian deviations by changing the temperature and density in order to affect $\chi$. This leads to incorrect fitting of the plasma parameters.  This can be illustrated following the above example with temperature. For a given $\xi\gtrsim1$, imaginary $\chi$ (Landau damping) is smaller for a super-Gaussian compared to a Maxwellian distribution, as shown in Fig.~\ref{fig:chi_plots}.  Consequently, the same temperature (as defined in Sec.~\ref{sec:vdfs}) will lead to narrower spectral peaks for the super-Gaussian.

\section{\label{sec:forward_models}Arbitrary Forward Model}

In this section we describe a forward model which maps a given set of discretized arbitrary VDFs to a corresponding TS spectrum, assuming that the VDFs are in quasi-equilibrium and that the plasma is unmagnetized~\cite{sheffield_2011}. The plasma physics scientific python package PlasmaPy~\cite{PlasmaPyv1.0} includes a built-in function which computes the TS spectrum given a set of Maxwellian plasma parameters (the ``Maxwellian forward model"), which includes ion and electron temperatures, densities, and drift velocities. We expand upon this code and develope a numerical function which computes the TS spectra for arbitrary VDFs, which we refer to as the ``arbitrary forward model". For a given VDF, the arbitrary forward model accepts an array of velocity values $\{v_j\}$ and corresponding VDF values $\{f_j\}$ which are meant to represent $f_j = f(v_j)$, where $f(v)$ is the true continuous VDF. The Maxwellian forward model in PlasmaPy was used as a benchmark for testing and confirming the validity of our arbitrary forward model.

\subsection{Numerical Implementation}

Evaluating the spectral density in Eq.~\ref{eq:spectral_density} requires computation of the integrals in Eqs.~\ref{eq:echi}-\ref{eq:ichi}, which includes integrating over a Landau contour to avoid the pole at $v = \omega/k$. We note that the susceptibility integrals can be recast into the form
\begin{equation}
    \chi(k, \omega) = \int_{-\infty}^{\infty} du\frac{g(u)}{\xi - u}
\end{equation}
with an appropriate change of variable $v\to u(v).$ This integral can be rewritten using Plemelj's formula\cite{sheffield_2011} as
\begin{equation}
    \chi(k, \omega) = i\pi g(\xi) + \text{p.v.}\int_{-\infty}^\infty du\frac{g(u)}{\xi - u},
    \label{eq:plemlj_chi}
\end{equation}
where $\text{p.v.}\int$ refers to the Cauchy principal value of the integral, defined as
\begin{multline}
    \text{p.v.}\int_{a}^b \frac{g(u)}{\xi - u}dx =\\ \lim_{\epsilon\to 0}\left[\int_{a}^{\xi-\epsilon} \frac{g(u)}{\xi - u}dx + \int_{\xi+\epsilon}^{b} \frac{g(u)}{\xi - u}dx \right].
\end{multline}
If we pick some small but finite standoff value $\epsilon = \phi$, we can calculate the principal value by standard numerical integration in the ranges $[a, \xi - \phi]$ and $[\xi+\phi, b]$, plus a correction term:
\begin{multline}
    \text{p.v.}\int_{a}^b \frac{g(u)}{\xi - u}dx \approx\\
    \left[\int_{a}^{\xi-\phi} \frac{g(u)}{\xi - u}dx + \int_{\xi+\phi}^{b} \frac{g(u)}{\xi - u}dx \right] + 2\phi g'(\xi).
\end{multline}
Using this result, the susceptibility can be approximated by
\begin{multline}
    \chi(k, \omega)\approx \int_{-\infty}^{\xi - \phi} du\frac{g(u)}{\xi - u} + \int_{\xi + \phi}^{\infty} du\frac{g(u)}{\xi - u}\\
    + i\pi g(\xi) + 2\phi g'(\xi).
    \label{eq:chi_approximation}
\end{multline}
The function $g(\xi)$ is proportional to $\partial f/\partial v$, which is numerically computed from the discretized input VDFs using a finite difference scheme to 4th order precision. The two definite integrals in Eq.~\ref{eq:chi_approximation} cross no poles, so they can be evaluated using a standard numerical integration scheme such as a Riemann sum. To minimize the number of integration points while maintaining precision, we implement a scheme in which the points are finely spaced close to the pole, where the integrand varies rapidly, but coarsely spaced away from the pole, where the integrand varies slowly.

\subsection{Numerical Errors}
\label{sec:numerical_errors}

The estimation of the $\chi$ integrals as done in Eq.~\ref{eq:chi_approximation} as part of the arbitrary forward model introduces some numerical error into the TS spectrum. Our algorithm uses the range of velocities defined for the input VDFs to determine an integration range. It also obtains the values of the integration points by interpolating from the input VDFs. This means that appropriate resolution of the input VDFs is necessary to minimize the error of the forward model. This constrains both the array spacing $\delta v$ of the input VDFs as well as the total range over which the input VDFs are defined.

For a VDF $f(v)$, the array spacing must be sufficiently small to resolve the features of the VDF and to precisely compute $\partial f / \partial v$. In general $\delta v$ should be less than a characteristic velocity associated with the resolution of $f(v)$, found by taking the ratio of the VDF with its derivative: $\delta v\leq f(v)/f'(v)$. The total range of the velocity array must be large enough so that the forward model includes all important features of the VDF. This can be expressed as $f(v)/\max(f)\ll 1$ for all $v$ outside the array range, where $\max(f)$ is the maximum value that $f(v)$ takes over all $v$. Both of these conditions are necessary for the numerical calculation of the spectral density. If the input VDF spacing is too sparse, then the $\chi$ integration may not have enough points to be accurate. If the VDF range is too small, then the finite integration is not a good approximation of the integral from $-\infty$ to $\infty$. Better quantification of the effects of the input velocity array on numerical integration will be the subject of future studies.

\begin{figure*}
    \centering
    \includegraphics[width = 1\linewidth]{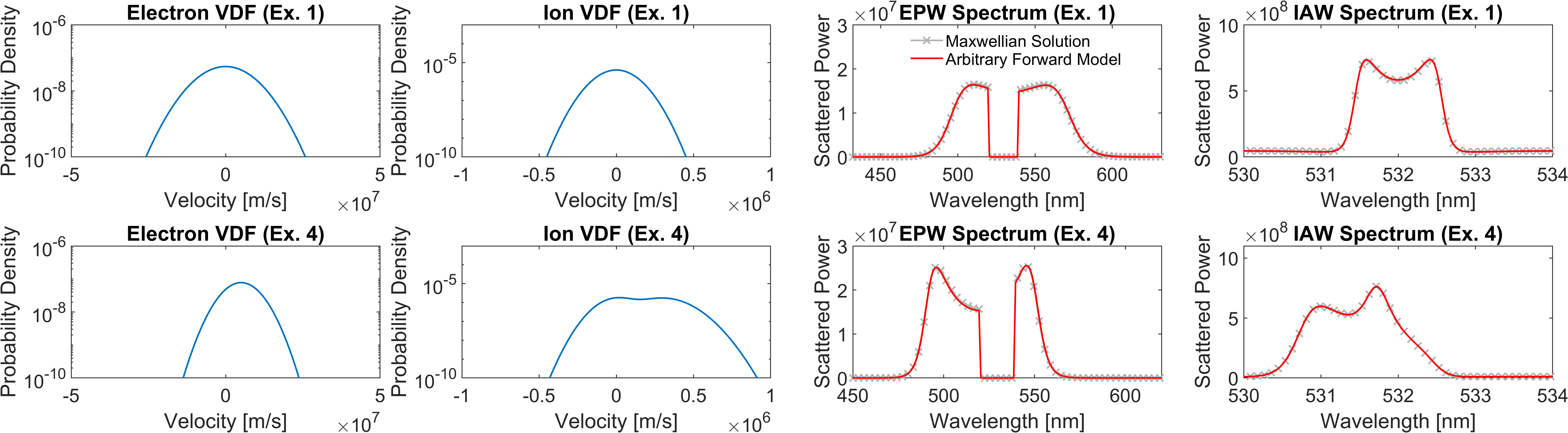}
    \caption{VDFs from a single Maxwellian (top row) or sums of Maxwellians (bottom row), and their corresponding TS spectra from both the Maxwellian (gray) and arbitrary (red) forward models. The plasma parameters are indicated in Table~\ref{tab:maxwellian_parameters} and indexed by the example numbers in the figure.
    }
    \label{fig:forwardmodel_benchmark}
\end{figure*}

We quantify the error in the Maxwellian case by comparing the output spectra generated by the arbitrary forward model using Maxwellian input VDFs to the corresponding output spectra from the Maxwellian forward model. Fig.~\ref{fig:forwardmodel_benchmark} compares the forward-modeled spectra for examples with VDFs that are Maxwellians or linear combinations of Maxwellians. There are a few small differences in the spectra, but they have good agreement overall, validating the arbitrary forward model. The pointwise relative error of the forward-modeled spectra from the arbitrary forward model was $\lesssim5\%$ for the examples shown in Fig.~\ref{fig:forwardmodel_benchmark}. In addition, those differences can be decreased arbitrarily by increasing the resolution of the $\chi$ integration scheme.

\begin{figure*}
    \centering
    \includegraphics[width = 1\textwidth]{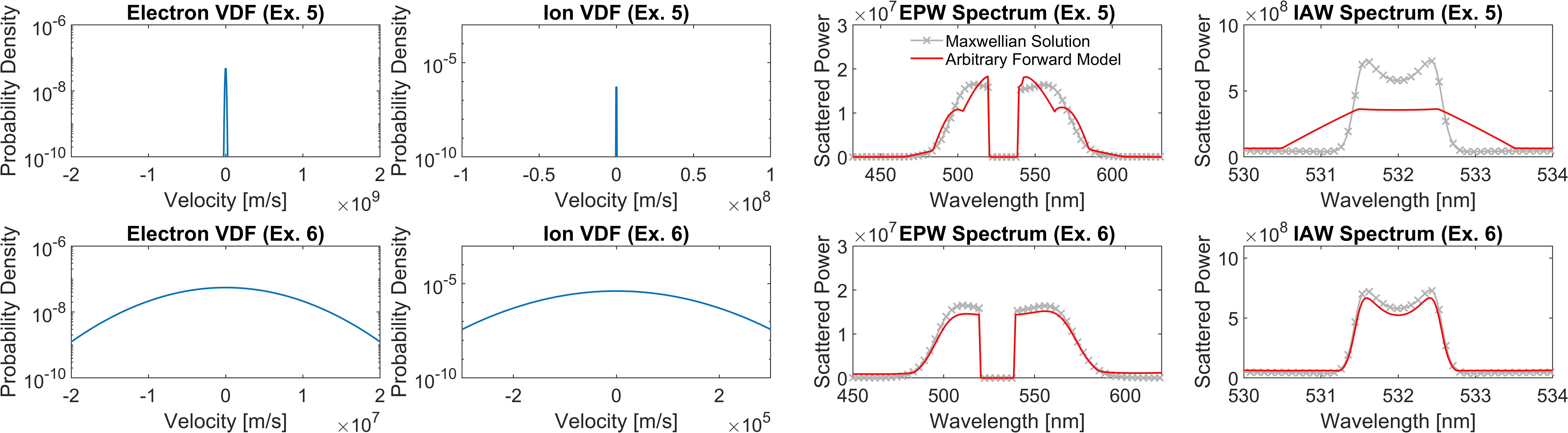}
    \caption{Arbitrary forward model applied to the same electron and proton VDFs at different velocity resolutions. (Top row) The velocity arrays that the VDFs are defined on are too sparse to properly resolve the VDF. (Bottom row) The velocity array does not cover a wide enough range of velocities to capture the entire VDF. For each, the corresponding TS spectra from both the Maxwellian (gray) and arbitrary (red) forward models are shown. The plasma parameters are listed in Table~\ref{tab:maxwellian_parameters}.}
    \label{fig:cold_hot_VDFs}
\end{figure*}

To illustrate the numerical effects of velocity resolution and VDF range, Fig.~\ref{fig:cold_hot_VDFs} shows the output spectra when the arbitrary forward model is applied to the same Maxwellian VDFs (Example 1 in Table~\ref{tab:maxwellian_parameters}), but where the velocity arrays on which the VDFs are defined have been modified. The ranges of the velocity arrays have been changed while leaving the array length fixed in both cases. In Example 5, the velocity range is much larger and the VDF is effectively defined by very few velocity bins. We refer to this as the ``narrow" case. In Example 6, the velocity range is smaller so that the tails of the VDFs are cut off. We refer to this as the ``wide" case. In the narrow case, the array spacing $\delta v$ is several orders of magnitude larger than the characteristic velocity $f(v)/f'(v)$ at some values of $v$, so the VDF is not well-resolved. In the wide case, the VDFs are wider than the velocity array. The VDFs are cut off at a factor of $\sim 0.01$ of their maxima, which is much larger compared to $\lesssim 10^{-10}$ in the well-resolved case shown in Fig.~\ref{fig:forwardmodel_benchmark}. In both cases, the resulting computed spectra deviate significantly from the Maxwellian forward model. 

\begin{figure}[h]
    \centering
    \includegraphics[width = \linewidth]{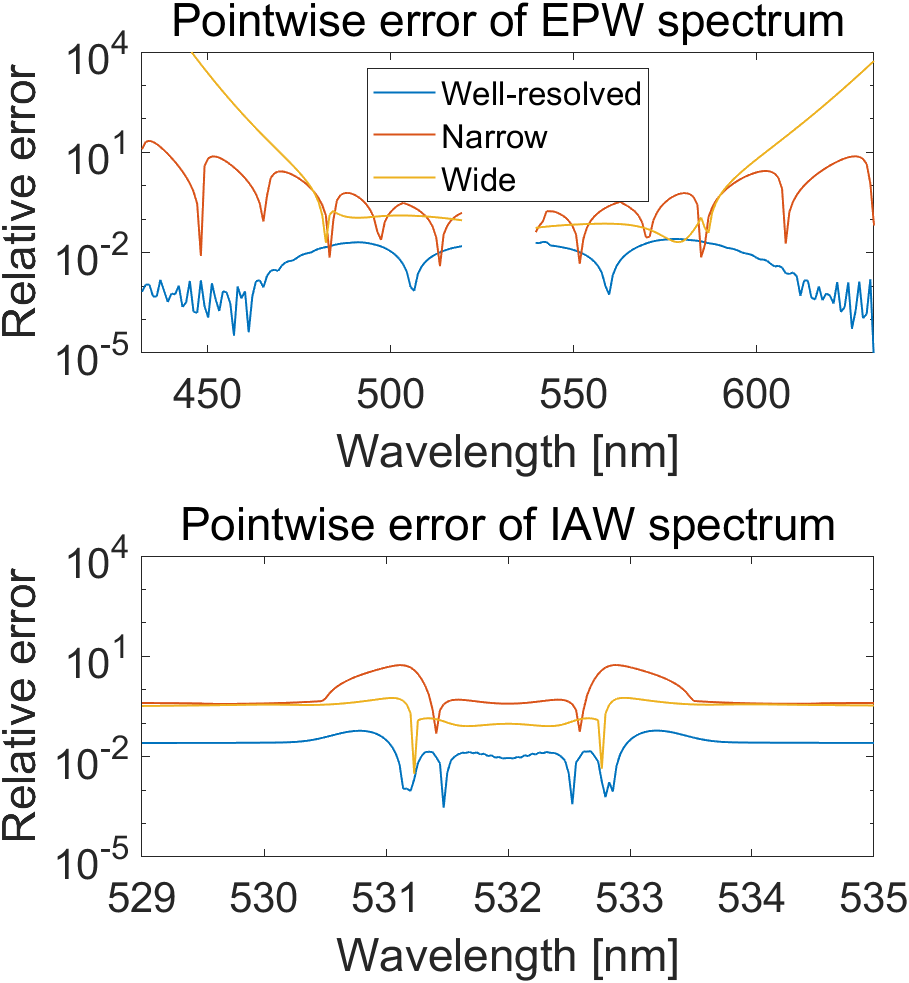}
    \caption{Pointwise relative error of the arbitrary forward model compared to the Maxwellian forward model for three different velocity resolutions. Both the narrow case, where the VDF is poorly resolved, and the wide case, where the VDF tails are cut off, result in significantly higher errors than the well-resolved case.}
    \label{fig:spectra_error}
\end{figure}

The pointwise relative error of the well-resolved, narrow, and wide cases are shown in Fig.~\ref{fig:spectra_error}. Due to the $\chi$ integration over the entire VDF, every point on the TS spectrum is affected by the entire VDF. Even in the wide case, where a subset of the VDF is very well-resolved, the pointwise error of the arbitrary forward model is worse across the entire TS spectrum.

\section{\label{sec:fitting_algorithms}Fitting Algorithms}

The current release of the plasma physics Python package PlasmaPy\cite{PlasmaPyv1.0} contains a built-in function which can fit Maxwellian plasma parameters to TS spectra using the DE algorithm from the Python package \code{lmfit}. This algorithm can also fit to sums of Maxwellians. Because the computation of the susceptibility function for Maxwellians is optimized by tabulation~\cite{poppe_1990}, using the Maxwellian forward model makes the fitting significantly faster than it would be with the arbitrary forward model. Therefore the main use case of the arbitrary forward model for fitting would be in fitting of TS spectra from non-Maxwellian VDFs.

There are several possible approaches to a non-Maxwellian iterative fitting algorithm, which produce fits to varying degrees of arbitrariness. If we had sufficient computing power, we could in theory treat each point on a discretized VDF array as its own parameter, and then fit to that set of parameters. However, implementing this approach with sufficient resolution of the velocity axis becomes computationally infeasible due to the large number of free parameters. Another potential approach would be to provide a number of pre-defined parametrized non-Maxwellian VDF models with low-dimension parameter spaces that the algorithm can fit to. This has the issue of being too restrictive, and a better method in this case would be to design individual forward models based around each of the VDF models. 

The approach our algorithm employs is to accept custom user-defined parametrized VDF models and to then fit an input TS spectra assuming that the VDFs are in the form of those user-defined models. This allows the user to fit to different non-Maxwellian models while still keeping the parameter space small. We compared the fits obtained by this approach with fits obtained from assuming Maxwellian VDFs to determine how well the arbitrary method performs.

As discussed in Sec.~\ref{sec:ts}, the EPW spectrum is negligibly affected by the ion VDFs, so the EPW spectrum is used first to fit the electron parameters while holding the ion parameters fixed at some arbitrary values that would be reasonable for the system. The IAW spectrum depends on both electron and ion parameters, so to reduce the time needed for fitting, the electron parameters can be fixed at the values obtained from fitting the EPW spectrum and the remaining ion parameters are fitted using the IAW spectrum. If necessary, this process could be further iterated with additional parameter constraints to refine the fit as needed.

\subsection{Procedure for comparing fitting algorithms}
We apply a general procedure to test both fitting algorithms and compare the results. First, we prepare synthetic electron and ion VDFs which have the same form as physically relevant VDFs. We then compute the EPW and IAW TS spectra of a plasma with these VDFs. Gaussian noise is added to make the resulting spectra more realistic, with a standard deviation given by $\sigma_n = 0.1\max(P(\lambda)),$ with $P(\lambda)$ being the TS spectrum. The VDFs are then reconstructed based on the fitted parameters and are compared to the initial input VDFs, while the parameters and their variances are compared to the input parameters. After obtaining the results from the fitting algorithms, an MCMC sampler is used to explore the parameter space and compute the one- and two- dimensional posterior probability distributions of the parameters in order to estimate the confidence in the fitted values.

After fitting the synthetic spectra, we analyze and compare the accuracy of the fits by calculating the $\chi^2$ statistic between the best-fit VDFs and the input VDFs, as well as the percent error of quantities such as the density, bulk flow velocity, and the equivalent temperature of the VDFs. We characterize the error bars associated with the fit and use these to search for correlations and degeneracies in the arbitrary forward model using an MCMC sampler~\cite{emcee2013}.

\begin{figure}[h]
    \centering
    \includegraphics[width=1\linewidth]{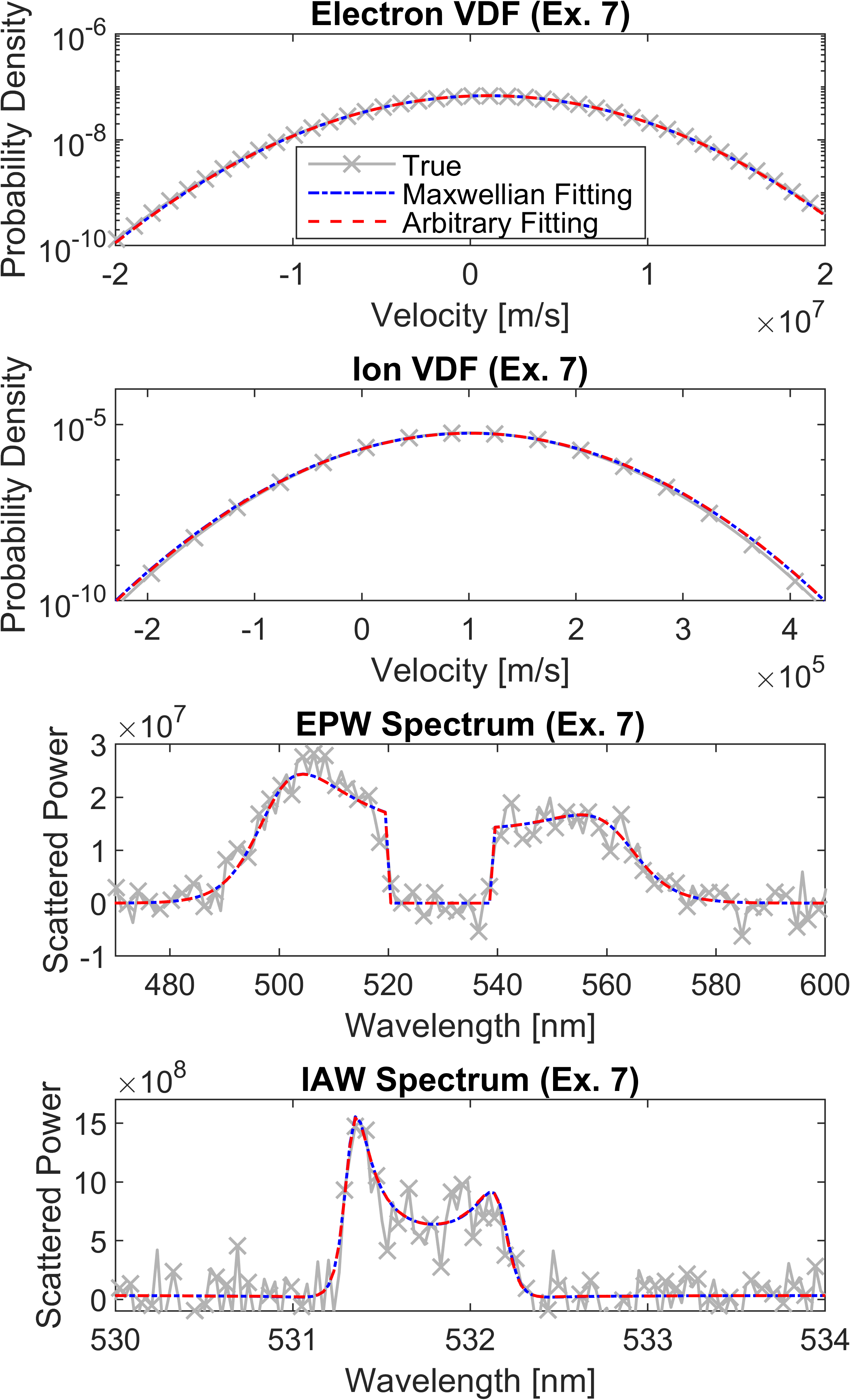}
    \caption{Synthetic TS spectrum derived from a Maxwellian electron VDF and Maxwellian proton VDF and fitted using both the arbitrary and Maxwellian fitting algorithms. The true VDFs which were used to generate the synthetic spectrum are shown in gray in the top row, and the resulting (notched) EPW and IAW spectra are shown in gray in the bottom row. The best-fit spectra and VDFs from the arbitrary fitting algorithm are shown in red and the best-fit spectra and VDFs from the Maxwellian fitting algorithm are shown in blue. The EPW and IAW spectra are scaled to have unit integrals when evaluated in SI units.}
    \label{fig:fitting_maxwellians}
\end{figure}

\subsection{Validating the arbitrary fitting algorithm}

\begin{table*}
    \centering
    \begin{tabular}{c|ccccccccc}
    Ex. \# & Ion species & $T_e$ [eV] & $v_e$ [m/s] & $T_i$ [eV] & $v_i$ [m/s] & $n_e$ [cm$^{-3}$]& $n_i$ [cm$^{-3}$] & $\kappa_C$ & $p_e$\\
    \hline
    7 & p & 200 & $10^6$ & 50 & $10^5$ & $4\times 10^{18}$ & $4\times 10^{18}$ & --- & ---\\

    8 & p & 300 & 0 & 50 & $10^5$ & $4\times 10^{18}$ & $4\times 10^{18}$ & --- & 3\\
    9 & (p, C$_{12}^{6+}$) & 200 & 0 & (100, 300) & $(2\times 10^5, 0)$ & $4\times 10^{18}$ & $(2.4, 0.267)\times 10^{18}$ & 2 & ---\\
    
\end{tabular}
    \caption{Plasma parameters used in the fitting examples in Figs.~\ref{fig:fitting_maxwellians}-\ref{fig:fitting_nonmaxwellian_2}. For Fig.~\ref{fig:fitting_nonmaxwellian_1}, the $p_e$ electron parameter follows the form of the 1D super-Gaussian in Eq.~\ref{eq:1d_supergaussian}, and for Fig.~\ref{fig:fitting_nonmaxwellian_2}, the $\kappa_C$ parameter for the carbon ions is as defined in Eq.~\ref{eq:kappa_def}.}
    \label{tab:fitting_example_parameters}
\end{table*}

We first fit a TS spectrum derived from Maxwellian VDFs in order to benchmark the arbitrary fitting algorithm against the existing fitting algorithm in PlasmaPy. For this test, the TS spectrum from Maxwellians is generated with the Maxwellian forward model and fitted with both the Maxwellian and arbitrary fitting models as shown in Fig.~\ref{fig:fitting_maxwellians}. The plasma parameters used to generate the VDFs are given in Table~\ref{tab:fitting_example_parameters}. The fitted VDFs agree with each other and with the input VDFs, validating the arbitrary fitting algorithm for fitting TS from Maxwellian VDFs. We can also verify this quantitatively by computing goodness-of-fit metrics for the fitted TS spectra and the VDFs, as shown in Table~\ref{tab:fitting_example_chisquare}. For the VDFs, we use the usual definition of $\chi^2$ goodness of fit:
\begin{equation}\label{eq:chi2_vdf}
    \chi^2_{VDF} = \frac{1}{N}\sum_N (F - I)^2,
\end{equation}
where the sum is over the indices of the discretized arrays for input function $I$ and fitted function $F$. In the case of the TS spectra, the Gaussian noise which is artificially added to the input spectra will increase $\chi^2$ on its own, so the usual definition of $\chi^2$ is normalized by $\sigma_n^2$ to account for this and allow comparison between the different fitting examples. Therefore for the TS spectra we use
\begin{equation}
    \chi^2_{TS} = \frac{1}{N\sigma_n^2}\sum_N (F - I)^2.
\end{equation}
Note that under this scheme it is still possible for a good fit to have $\chi^2<1,$ if the randomly generated synthetic noise happens to contribute less than the expected squared-error of $\sigma_n^2.$ From Table~\ref{tab:fitting_example_chisquare} we see that the Maxwellian and arbitrary fitting algorithms are approximately on par with each other when fitting Maxwellians.

\section{Results}\label{sec:results}
\subsection{Comparisons between best-fit VDFs}

\begin{figure}
    \centering
    \includegraphics[width =1\linewidth]{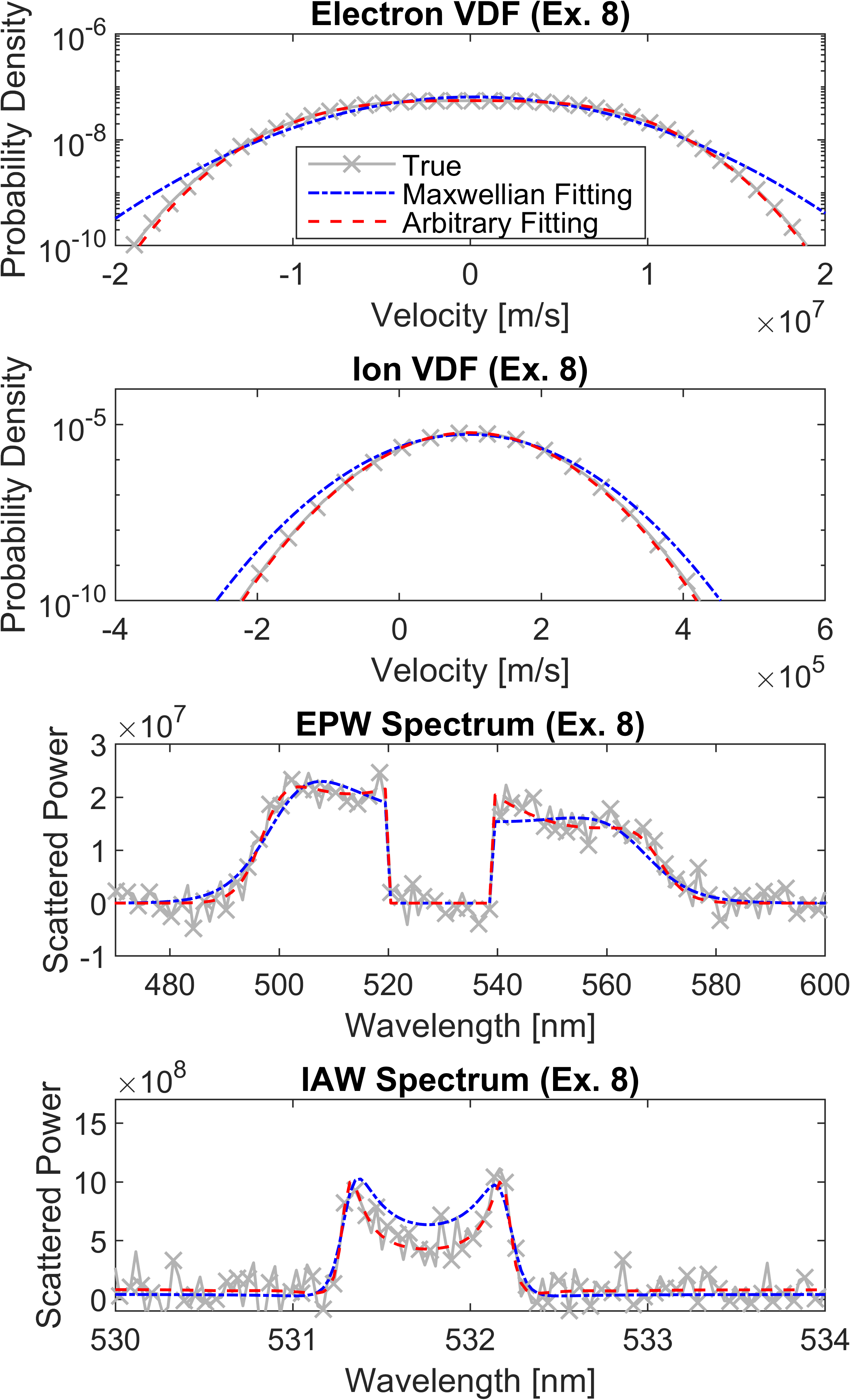}
    \caption{TS spectrum from non-Maxwellian (super-Gaussian) electrons and Maxwellian protons fitted using the arbitrary and Maxwellian fitting algorithms. The same plot organization and color scheme are used as in Fig.~\ref{fig:fitting_maxwellians}.}
    \label{fig:fitting_nonmaxwellian_1}
\end{figure}
\begin{figure}
    \centering
    \includegraphics[width =1\linewidth]{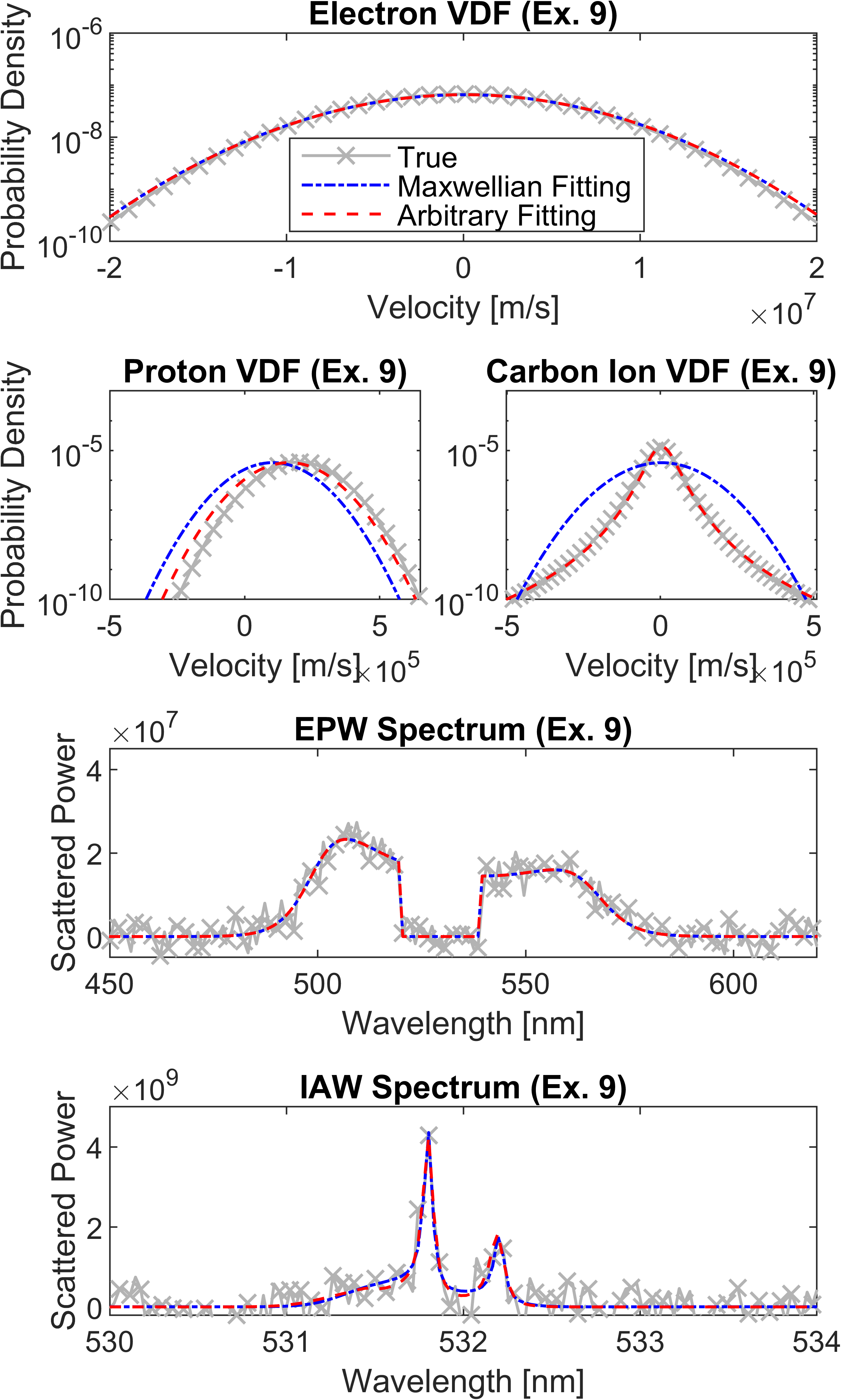}
    \caption{TS spectrum from Maxwellian electrons, Maxwellian protons, and non-Maxwellian (kappa-distributed) carbon ions fitted using the arbitrary and Maxwellian fitting algorithms. The same plot organization and color scheme are used as in Fig.~\ref{fig:fitting_maxwellians}, except the second row now shows an additional figure for the carbon ion VDF.}
    \label{fig:fitting_nonmaxwellian_2}
\end{figure}

After validating the arbitrary fitting algorithm, we test it on synthetic TS spectra from plasmas with non-Maxwellian VDFs which are computed using the arbitrary forward model. The resulting fits are compared to those generated by applying the Maxwellian fitting algorithm to the same data. The plasma parameters for these non-Maxwellian parameters are also given in Table~\ref{tab:fitting_example_parameters}.

\begin{table*}
    \centering
    \begin{tabular}{c|cccc|cccc}
    \multicolumn{9}{c}{}\\
    & \multicolumn{4}{c|}{Maxwellian Fitting Algorithm} & \multicolumn{4}{c}{Arbitrary Fitting Algorithm}\\
    
    Ex. \# & $\chi^2_{\text{EPW}}$ & $\chi^2_{\text{eVDF}}$ & $\chi^2_{\text{IAW}}$ & $\chi^2_{\text{iVDF}}$ & $\chi^2_{\text{EPW}}$ & $\chi^2_{\text{eVDF}}$ & $\chi^2_{\text{IAW}}$ & $\chi^2_{\text{iVDF}}$\\
    \hline
    7 & 1.005 & $2.97\times 10^{-20}$ & 1.203 & $4.87\times 10^{-16}$ & 1.006 & $3.58\times 10^{-20}$ & 1.200 & $6.19\times 10^{-16}$\\

    8 & 1.195 & $6.92\times 10^{-18}$ & 1.765 & $7.29\times 10^{-15}$ & 0.936 & $5.34\times 10^{-20}$ & 1.121 & $2.475\times 10^{-16}$\\

    9 & 1.037 & $3.76\times 10^{-19}$ & 0.915 & $(0.29, 1.75)\times 10^{-12}$ & 1.036 & $3.73\times 10^{-19}$ & 0.893 & $(4.57, 5.19)\times 10^{-14}$\\
    
\end{tabular}
    \caption{Values of the $\chi^2$ goodness of fit parameter for the different fitted TS spectra and VDFs for both fitting algorithms. The $\chi^2$ is calculated separately for the EPW spectrum, IAW spectrum, electron VDF (eVDF), and ion VDFs (iVDF). The listed values for the VDFs are given by the usual definition of $\chi^2$ (Eqn.~\ref{eq:chi2_vdf}). The listed values for the TS spectra are normalized by the synthetic noise and also have the noise subtracted out. The arbitrary fitting algorithm performs at least as well as the Maxwellian fitting algorithm in all cases shown here.}
    \label{tab:fitting_example_chisquare}
\end{table*}

First, we study the combination of a non-thermal 1D super-Gaussian distribution for the electron VDF and a drifting Maxwellian distribution for the ion VDF.  As can be seen in Fig.~\ref{fig:fitting_nonmaxwellian_1}, the electron VDFs are not well-fit by a Maxwellian. Additionally, the moments of the VDF, which are the electron density and the temperature, are not accurately recovered. 

The algorithm attempts to fit a Maxwellian to the flattened top of the super-Gaussian, but greatly over-estimates the electron temperature in doing so. Additionally, while the ions are Maxwellian, the ion parameters are still not fit correctly with the Maxwellian fitting algorithm.  This is because the IAW spectrum has non-trivial dependence on the electron VDF, so that the errors in the electron VDF effectively propagate into the IAW fitting. We also see this in Table~\ref{tab:fitting_example_chisquare}, as the $\chi_{IAW}^2$ and $\chi_{iVDF}^2$ values are significantly higher for the Maxwellian fitting algorithm than for the arbitrary fitting algorithm. For the arbitrary fitting algorithm, because the correct models are used for both the electrons and the ions, the VDFs are both well-fit. However, if the wrong VDF model were to be used for the electrons, the arbitrary fitting algorithm could suffer from the same error propagation issue. This emphasizes the importance of using the correct VDF models for fitting. 

Second, we study the synthetic spectrum of a drifting Maxwellian electron VDF, a drifting Maxwellian proton VDF, and a kappa-distributed ion (C$^{+6}$) VDF. The relative densities of the protons and carbon ions are chosen to make the plasma quasi-neutral. In Fig.~\ref{fig:fitting_nonmaxwellian_2}, we see that electron VDF is well-fit by the Maxwellian fitting algorithm, but the proton and carbon VDFs are fitted inaccurately. Similarly to the previous example where the poorly-fit non-Maxwellian electron VDF impacted the proton VDF fitting, here the non-Maxwellian carbon VDF causes a similar effect. 

Although the best-fit ion VDFs for the Maxwellian fitting algorithm are quite different from the corresponding best-fit VDFs from the arbitrary fitting algorithm, we see that the resulting best-fit IAW spectra look sufficiently similar to be within noise effects of each other. This can also be seen in Table~\ref{tab:fitting_example_chisquare}, as the $\chi_{IAW}^2$ values are similar for the two fitting algorithms, but the $\chi_{iVDF}^2$ values are significantly higher for the Maxwellian fitting algorithm. These results suggest a near-degeneracy in the TS forward model. 

The existence of near-degeneracies in the TS forward model raises the risk of getting a good fit on the TS spectrum which corresponds to an inaccurate VDF. This can be mitigated by restricting the fitting algorithm to a physically motivated VDF model and using other diagnostic results to put limits on the range of plasma parameters. The following section discusses how to identify degeneracies arising from different VDF models.

\subsection{\label{sec:error_bars}Uncertainty Analysis}

After finding the best-fit plasma parameters, we analyze the robustness and uncertainty in the fits. This allows us to determine which aspects of the fits we can be confident in, as well as reveal TS spectrum degeneracies which could make the fits inaccurate. The Python module \code{emcee}~\cite{emcee2013} was used to estimate the posterior probability distributions (PPDs) of the fits using an MCMC sampling algorithm, which is given initial values at the best-fit values from the arbitrary fitting algorithm. Using the PPDs we estimate error bars for each fitted parameter and show that when fitting TS spectra from non-Maxwellian distributions, fits using the arbitrary fitting algorithm derived plasma parameters with greater accuracy than the Maxwellian fitting algorithm. 

\begin{figure*}
    \centering
    \includegraphics[width=1\linewidth]{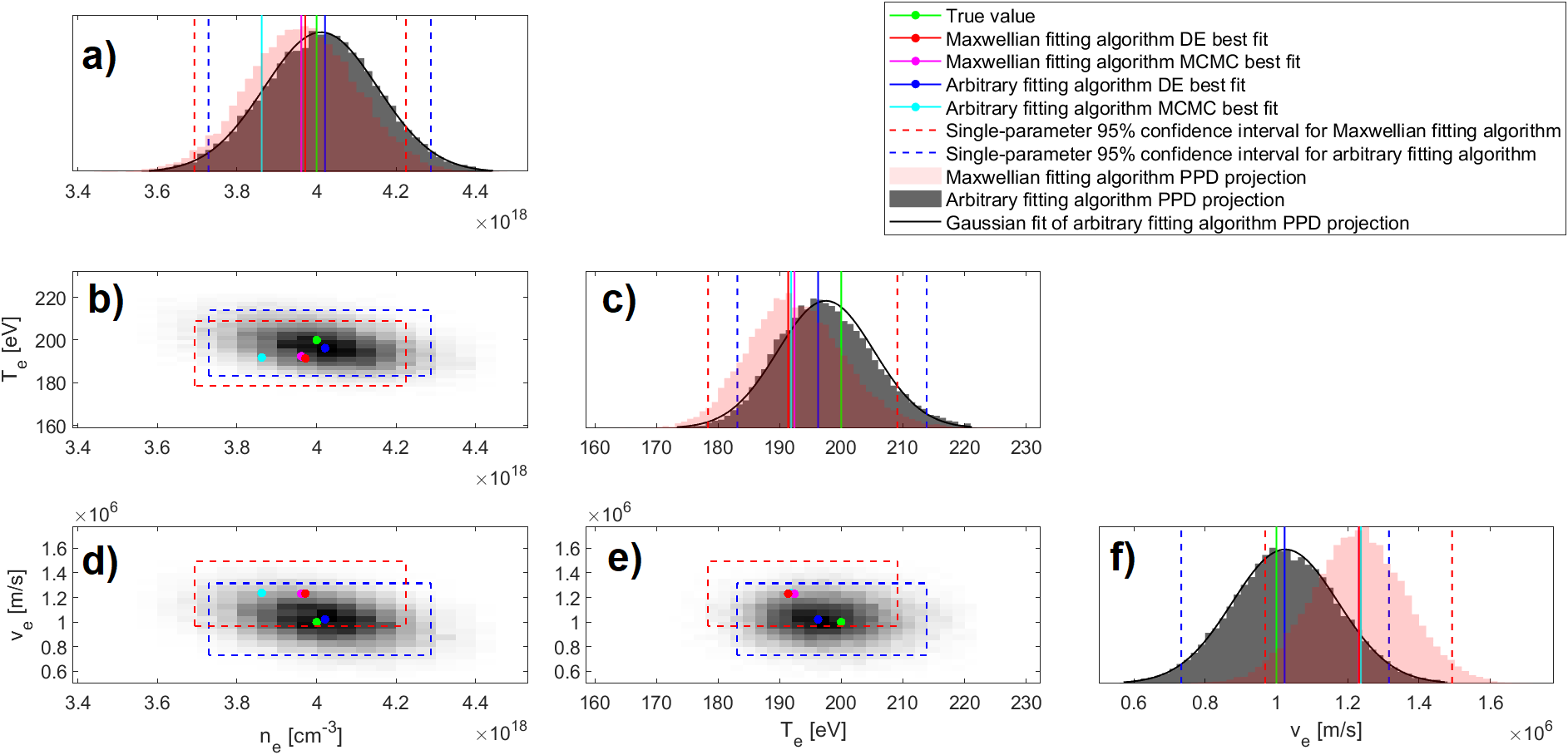}
    \caption{1D and 2D projections of the PPDs associated with applying the arbitrary and Maxwellian fitting algorithms to a TS spectrum from Maxwellian VDFs. The plots labeled a), c), and f) are the 1D projections. The plots labeled b), d), and e) are the 2D projections. The green dots and solid lines mark the true values of the plasma parameters. The red dots and solid lines indicate the best-fit values obtained from the arbitrary fitting algorithm, and the blue dots and lines are the best-fit parameters from the Maxwellian fitting algorithm. The PPD projections from the arbitrary and Maxwellian fitting algorithms are in gray and light red, respectively, and used to estimate 95\% confidence intervals for the fitted parameters in both algorithms, which are the dashed red and blue lines. The 1D PPD projections for the arbitrary fitting algorithm are fitted by Gaussians, shown by the solid black curves. For comparison, the parameters at which the PPDs are maximized are given by the magenta dots and solid lines for the Maxwellian fitting algorithm, and the cyan dots and solid lines for the arbitrary fitting algorithm.}
    \label{fig:maxwellian_corner}
\end{figure*}

\begin{figure*}
    \centering
    \includegraphics[width=1\linewidth]{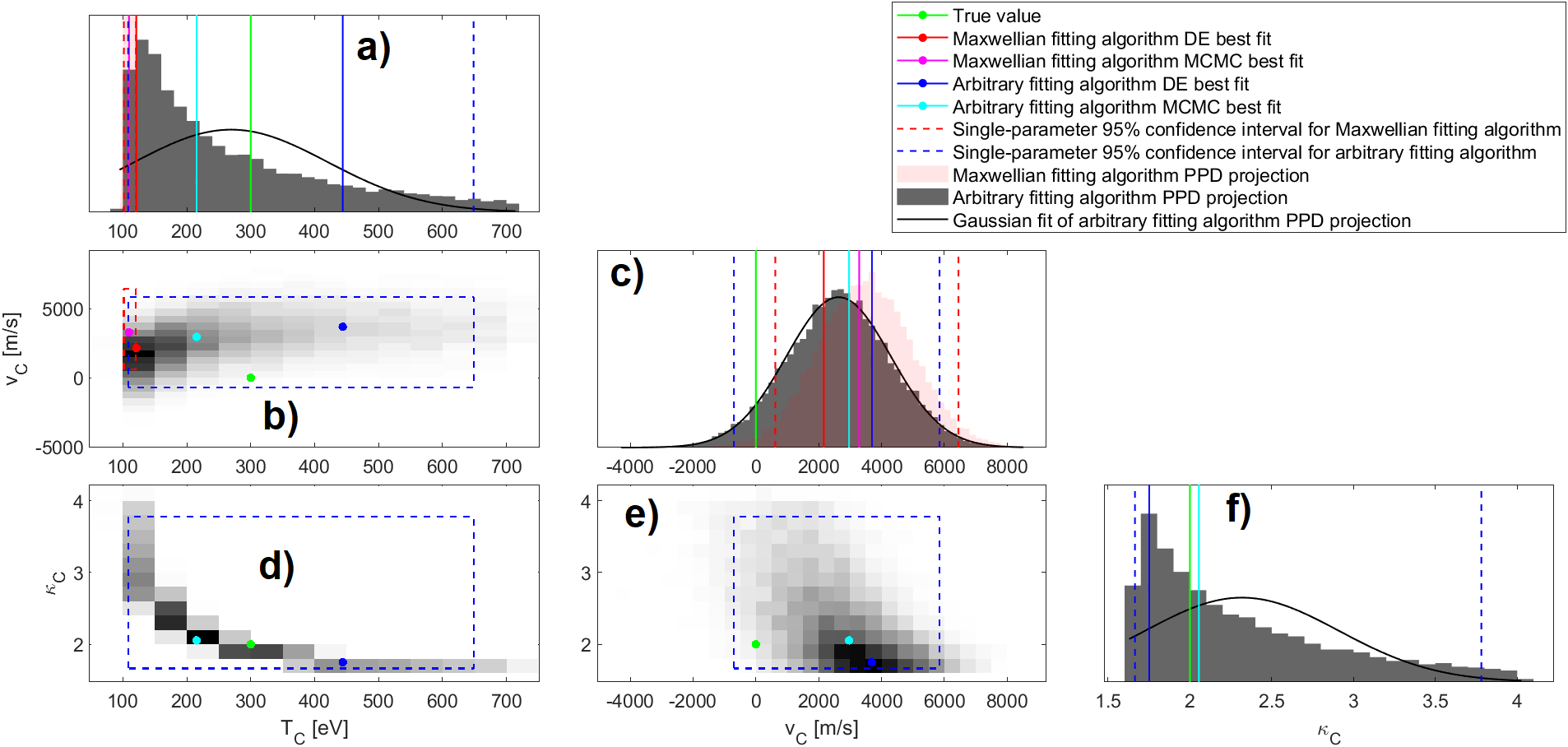}
    \caption{1D and 2D projections of the PPDs associated with applying the arbitrary and Maxwellian fitting algorithms to an IAW spectrum from a kappa-distributed ion VDF. The same color scheme is used as in Fig.~\ref{fig:maxwellian_corner}. Note that the $\kappa$ parameter is not fitted by the Maxwellian fitting algorithm, so there are no red/magenta elements in the plots associated with the $\kappa$ parameter.}
    \label{fig:kappa_corner}
\end{figure*}

Fig.~\ref{fig:maxwellian_corner} shows the PPD associated with the fitting of the EPW spectrum in Fig.~\ref{fig:fitting_maxwellians}. We see that the 1D PPD projections (Figs.~\ref{fig:fitting_maxwellians}a,c,f) approximately follow Gaussian distributions. The two-dimensional slices (Figs.~\ref{fig:fitting_maxwellians}b,d,e) also look approximately symmetric, indicating that the variables are effectively fitted independently of each other. From this we conclude that the fitted parameters which were output by the arbitrary fitting algorithm are accurate, as there is one clear location in parameter space where the posterior probability density reaches a maximum, and all marginal distributions also reach their maxima. The $95\%$ confidence intervals (dashed lines) for each parameter are shown for reference. 

Both fitting algorithms produce approximately equal-sized uncertainties for each parameter, and the true values are contained within these uncertainties. This benchmarks the arbitrary fitting algorithm to be approximately on par with the Maxwellian fitting algorithm when fitting TS spectra from Maxwellians. Some of the best-fit values in Fig.~\ref{fig:maxwellian_corner} visually appear far away from the true parameters, but this is due to a rescaling of the relevant axes in the figure in order to see the PPD features, and does not represent the actual ranges of parameters over which the algorithm was allowed to explore (this is especially the case with the velocity parameter, which was fit over a large range). For instance, the fitting algorithm used to produce the results of Fig.~\ref{fig:maxwellian_corner} ranged over electron temperatures of $10$--$2000\;\mathrm{eV},$ drift velocities of $-10^{7}$--$10^{7}\;\mathrm{m/s},$ and electron densities of $10^{17}$--$10^{19}\;\mathrm{cm^{-3}}.$  When taking the full ranges into account, the errors of the best fit parameters are on the order of $\sim 1\%$ of the total parameter range, which shows that the algorithms find the correct plasma parameters even when allowed to explore over a very large region of parameter space.

Fig.~\ref{fig:kappa_corner} shows the PPD corresponding to the fitting of the C$^{6+}$ ion parameters from the IAW spectrum in Fig.~\ref{fig:fitting_nonmaxwellian_1} using the arbitrary fitting algorithm. In order to focus on the kappa-distributed carbon ion VDF parameters, the electron and proton parameters were all held fixed at best-fit values during the MCMC sampling. Unlike the Maxwellian case, the 1D marginal distributions (Figs.~\ref{fig:kappa_corner}a,c,f) are no longer Gaussian and the two-dimensional slices (Figs.~\ref{fig:kappa_corner}b,d,e) illustrate a strong correlation between the fitted electron temperature and the spectral index $\kappa$. This is because both the temperature and $\kappa$ affect the width of the VDF, so they are degenerate. We see that the 95\% confidence intervals for $T$ and $\kappa$ are large, which in the absence of other information might suggest that the fit is not robust. However, from the full PPD we see that the single-parameter confidence intervals are misleading due to the correlated parameters. It is possible that if the initial guesses for the DE fitting algorithm are chosen poorly, the algorithm would land somewhere along the curve where the value of the PPD is high, but far from the true values, therefore resulting in inaccurate fitted parameters. In practice, physical constraints obtained by analyzing other experimental data must be applied as boundaries on the fitting parameters to break this degeneracy. 

We also see that the Maxwellian-fitted temperature is significantly lower than the true equivalent temperature. This can be explained by the fact that at the location of the carbon spectral peaks at $\xi \approx 1.57$ in this example, the imaginary (Landau damping) term of $\chi$ for a kappa distribution is less than that of a Maxwellian (see Fig.~\ref{fig:chi_plots}), which the fitting algorithm compensates for by lowering the Maxwellian temperature. It is important to note that while the Maxwellian Landau damping term remains above the corresponding term for super-Gaussians at high $\xi$ (corresponding to wavelengths far from the incident laser wavelength, see Fig.~\ref{fig:chi_plots}), the Maxwellian Landau damping term actually drops below the kappa Landau damping term at high $\xi$. This indicates that in a parameter regime where the VDF is a kappa distribution but the spectral peaks lie at high $\xi$, the error in the Maxwellian fitting could be reversed, which further illustrates the problems with applying the Maxwellian fitting algorithm to TS from non-Maxwellian plasmas. 

Although the arbitrary fitting algorithm can perform better than the Maxwellian fitting algorithm in fitting TS spectra from non-Maxwellian VDFs, we have shown that the presence of non-Maxwellian VDF models can make the results of the fitting difficult to interpret without taking into account the details of how the different parameters of the non-Maxwellian VDF model correlate and affect the forward model. The use of the MCMC sampler on a synthetic distribution is one means by which we can study these correlations for given VDF models. 

\section{Conclusions\label{sec:conclusions}}

We have developed an arbitrary forward model which can compute integral-normalized TS spectra from arbitrary discretized electron and ion VDFs. The arbitrary forward model is benchmarked against a forward model which uses a tabulated plasma dispersion relation to compute the TS spectra for VDFs which are Maxwellian or linear combinations of Maxwellians. We discuss the numerical error associated with the arbitrary forward model and provide examples of how to minimize these effects.

The arbitrary forward model is used to implement an iterative fitting algorithm that accepts TS spectra as inputs and recovers plasma parameters defined by arbitrary user-defined VDF models. We show that the arbitrary fitting algorithm performs similarly to a Maxwellian TS fitting algorithm for Maxwellian VDFs, but outperforms the Maxwellian fitting algorithm for non-Maxwellian VDFs. The arbitrary forward model can be used to fit the plasma parameters associated with any parameterizable non-Maxwellian VDF model which satisfies the basic TS assumptions of quasi-neutrality and an unmagnetized plasma (although this assumption could be relaxed with a suitable extension of the TS analytic model).

The fitting algorithm can be run with an MCMC sampler to estimate the PPD over the parameter space. This enables us to estimate the uncertainties of the fitted parameters and show that the numerical errors associated with the arbitrary forward model do not have a significant impact on the accuracy of the fitting. In the case of kappa VDFs, we also determine from the PPD that several parameters are linearly-dependent (i.e. degenerate). Use of the arbitrary fitting algorithm to fit synthetic spectra from other non-Maxwellian VDF models could also reveal correlated parameters in those models. Further work is needed to gain a better understanding of which VDF models contain these correlations and methods by which we can best mitigate them. Each VDF model could be examined separately using the tools we have developed in order to determine what parameter correlations are relevant in the PPDs. In addition, the computationally-heavy calculation of $\chi$ and the resulting runtime increase could make the arbitrary forward model inconvenient to use in fitting experimental data. Further work should be done in either optimizing the speed of the forward model or in designing other forward model schemes. For instance, forward models which are tailored for specific non-Maxwellian VDF models could be developed and benchmarked against our arbitrary forward model. Although these types of forward models would only be suited for those particular VDF models and thus inherently less generalized, it is possible that additional speed optimizations can be made in these cases. 

\begin{acknowledgments} 
We thank R. Follett for many valuable discussions related to this work. This work was supported by the U.S. Department of Energy (DOE) National Nuclear Security Administration (NNSA) under Award Nos. DE-NA0004033, DE-NA0003856, and DE-SC0020431, the University of Rochester, and the New York State Energy Research and Development Authority. This work was also supported by NASA under Grant No. 80NSSC19K0493.

This report was prepared as an account of work sponsored by an agency of the U.S. Government. Neither the U.S. Government nor any agency thereof, nor any of their employees, makes any warranty, express or implied, or assumes any legal liability or responsibility for the accuracy, completeness, or usefulness of any information, apparatus, product, or process disclosed, or represents that its use would not infringe privately owned rights. Reference herein to any specific commercial product, process, or service by trade name, trademark, manufacturer, or otherwise does not necessarily constitute or imply its endorsement, recommendation, or favoring by the U.S. Government or any agency thereof. The views and opinions of authors expressed herein do not necessarily state or reflect those of the U.S. Government or any agency thereof.

This research made use of PlasmaPy version 2023.1.0, a community-developed open source Python package for plasma research and education (PlasmaPy Community et al. 2023).

\end{acknowledgments}

\bibliography{aipsamp.bib}

\end{document}